\documentstyle[12pt,amsmath,amssymb,review]{elsarticle}

\newcommand{\bfa}[1]{{\mathbf{#1}}}
\newcommand{\bfb}[1]{{\mbox{\boldmath $#1$}}}

\newcommand{\rhob}{\bfb{\rho}}

\newcommand{\sigmab}{\bfb{\sigma}}
\newcommand{\Sigmab}{\bfb{\Sigma}}

\newcommand{\xb}{\bfa{x}}

\newcommand{\Lb}{\bfa{L}}

\newcommand{\Rb}{\bfa{R}}

\newtheorem{thm}{Theorem}
\newtheorem{lem}[thm]{Lemma}
 
\newproof{pol}{Proof of Equation (\ref{eq:Rhob_jl})} 
\newproof{pol2}{Proof of Theorem~\ref{thm:CholSums}}
\title{Direct formulation to Cholesky decomposition of a general nonsingular correlation matrix.\footnote{The work described in this paper was funded in part by the US National Institute of Mental Health (RC2 MH089951, principal investigator P.F.S.) as part of the American Recovery and Reinvestment Act of 2009.}}
\author{Vered Madar\footnote{Email address for correspondence: vered.madar@gmail.com} \\
University of North Carolina at Chapel Hill, North Carolina, USA.
}

\begin{document}

\begin{abstract}
We present two novel and explicit parametrizations of Cholesky factor of a nonsingular correlation matrix.
One that uses semi-partial correlation coefficients, and a second that utilizes differences between the successive ratios of two determinants. 
To each, we offer a useful application.
\end{abstract}

\maketitle

\section{Cholesky decomposition - Introduction}
For a positive-definite symmetric matrix Cholesky decomposition provides a unique representation in the form of $\Lb\Lb^T$, 
with a lower triangular matrix $\Lb$ and the upper triangular $\Lb^T$. 
Offered by a convenient $O(n^3)$ algorithm, 
Cholesky decomposition is favored by many for expressing the covariance matrix~\cite{Pourahmadi2011}. 
 The matrix $\Lb$ itself can be used to transform independent normal variables into dependent multinormal~\cite{Moonan57} which is particularly useful for Monte Carlo simulations. 

Explicit forms of $\Lb$ are known for limited correlation structures such as the equicorrelated~\citep[pp. 104]{Tong1990}, tridiagonal~\cite{MiwaHayterKuriki2003}, and the multinomial~\cite{TanabeSagae1992}. The general correlated case is typically computed by using spherical parametrizations~\cite{PinheiroBates1996,RapisardaBrigoMercurio2007,RebonatoJackel2007,MittelbachMatthiesenJorswieck2012}, a multiplicative ensemble of trigonometric functions of the angles between pairs of vectors. Others may use Cholesky matrix~\cite[pp. 49]{CookeEtAl2011} that utilizes the multiplication of partial correlations. 
 
In this paper, we will present two explicit parametrizations of Cholesky factor for a positive-definite correlation matrix. 
Both parametrizations offer a preferable, simpler alternatives to the multiplicative forms of spherical parametrization and partial correlations. 
In Section~\ref{sec:1stL} we show that the nonzero elements of Cholesky factor are the \textit{semi-partial correlation coefficients}
\[
  \rhob_{ij(1,\ldots,i-1)} 
= \frac{\rho_{ij} - \rhob_i^{*j} \Rb_{i-1}^{-1} \rhob_i}
          {\sqrt{1 - \rhob_i \Rb_{i-1}^{-1} \rhob_i^T}} ,
\]
where $\Rb_{i-1}^{-1}$ is the inverse of the correlation matrix $\Rb_i = (\rho_{kj})_{k,j=1}^{i-1}$, $\rhob_i^{*j} = (\rho_{1j}, \rho_{2j}, \ldots, \rho_{i-1,j})$ and $\rhob_i = \rhob_i^{*i}$. The order of the $\rhob_{ij(1,\ldots,i-1)}$s is determined by Cholesky factorization, and the notations are borrowed from Huber's trivariate discussion of semi-partial correlation in regression~\cite{Huber1981}.
 In Section~\ref{sec:2ndL} we uncover that the squares, $\rhob_{ij(1,\ldots,i-1)}^2$, are equivalent to the differences between two successive ratios of determinants, 
and we use this equivalence to construct the second parametrization for $\Lb$. In Section~\ref{sec:gencov} we extend the representation of $\Lb$ to the structure of a covariance matrix, 
and in Section~\ref{sec:welldefL} we study two inequality conditions that are essential for the positive-definiteness of $\Lb\Lb^T$.
We conclude this paper by offering two possible applications, one for each of the suggested forms. 
In Section~\ref{sec:ttest} we present a simple $t$-test that employs the semi-partial correlation structure for testing the dependence of a single variable upon a set of multivariate normals. 
In Section~\ref{sec:randomcorr} we utilize the second parametrization to design a simple algorithm for the generation of random positive-definite correlation matrices. 
We end the paper with the simple case of generalization of random AR(1) correlation in Section~\ref{sec:AR1}.

\section{The first parametrization for Cholesky factor}\label{sec:1stL}
Let $\Rb_n = (\rho_{ij})_{ij=1}^n$ be a positive-definite correlation matrix, for which each sub-matrix $\Rb_k = (\rho_{ij})_{ij=1}^k$ is positive-definite. 
Let also $\Lb = (l_{ij})_{ij = 1}^n$ be Cholesky factor of $\Rb$, 
$|\Rb|$ be the determinant of $\Rb$, $\Rb^{-1}$ its inverse, and $\rhob_i^{*j} = (\rho_{1j}, \rho_{2j}, \ldots, \rho_{i-1,j})$ for $j\ge i$, so $\rhob_i \equiv \rhob_i^{*i}$. 
 To simplify writing also set $\Rb_0^{-1} \equiv 1$. 
 The first representation of $\Lb$ will use 
 the semi-partial correlations $l_{ji} = \rhob_{ij(1,\ldots,i-1)} = \frac{\rho_{ij}-\rhob_i^{*j}\Rb_{i-1}^{-1}\rhob_i^T}{\sqrt{1 - \rhob_i \Rb_{i-1}^{-1} \rhob_i^T}}$, $i\le j$,
\begin{equation}\label{chol:L1}
     \Lb = \left( \begin{array}{ccccc}
       1 & 0 & 0 & \cdots & 0 \cr
       \rho_{12} & \sqrt{1 - \rho_{12}^2} & 0 
& \cdots & 0\cr 
       \rho_{13} & \frac{\rho_{23} - \rho_{12}\rho_{13}}{\sqrt{ 1 - \rho_{12}^2}} & 
         \sqrt{1 -  \rhob_3 \Rb_2^{-1} \rhob_3^T} & \cdots & 0\cr  
\vdots & \vdots & \vdots 
& \ddots & 0 \cr
\rho_{1n} &  
     \frac{\rho_{2n} - \rho_{12}\rho_{1n}}{\sqrt{1 - \rho_{12}^2}} &  
           \frac{\rho_{3n} - \rhob_3^{*n} \Rb_2^{-1} \rhob_3^T}{\sqrt{1 -  \rhob_3 \Rb_2^{-1} \rhob_3^T}} & 
      \cdots & \sqrt{1 -  \rhob_n \Rb_{n-1}^{-1} \rhob_n^T} 
\end{array}\right)
\end{equation}
For $i=j$, we have $\rho_{ii(1,\ldots,i-1)} = \sqrt{1-\rhob_i \Rb_{i-1}^{-1} \rhob_i^T}$, and $1 - \rhob_i \Rb_{i-1}^{-1} \rhob_i^T > 0$ for positive-definite $\Rb_i$. 
Some may recognize $1 - \rhob_i \Rb_{i-1}^{-1} \rhob_i^T$ as the \textit{schur-complement} 
of the matrix $\Rb_{i-1}$ inside $\Rb_i$ from the formula for computing the determinant of $\Rb_i$, using the block matrix $\Rb_{i-1}$~\citep[pp. 188]{Harville1997},
\begin{equation}\label{eq:schur_det}
   |\Rb_i|  =  \left|
 \begin{array}{cc} \Rb_{i-1} & 
 \rhob_i^T \\ \rhob_i & 1\end{array}
  \right| 
   = |\Rb_{i-1}| (1-\rhob_i \Rb_{i-1}^{-1} \rhob_i^T).
\end{equation} 
 
To show that $\Rb_n = \Lb\Lb^T$ we introduce Theorem~\ref{thm:CholSums}.  
\begin{thm}\label{thm:CholSums}
For $i \ge 1$ and $n \ge j \ge i+1$,
\begin{equation}\label{Q_recursive}
    \rhob_{i+1}^{*j}\Rb_i^{-1} \rhob_{i+1}^T  
           = \sum_{k=1}^i \rho_{ki(1,\ldots ,k-1)} \cdot \rho_{kj(1,\ldots ,k-1)} . 
\end{equation}
\end{thm} 
By the virtue that Cholesky factor of a positive-definite matrix has a unique representation, Theorem~\ref{thm:CholSums} will serve as a general proof for the form (\ref{chol:L1}). 
Some may recognize the Eq.~(\ref{Q_recursive}) in Theorem~\ref{thm:CholSums} as the inner-product used for the familiar algorithm of Cholesky Decomposition~\citep[pp. 235]{Harville1997}:
\[ 
   l_{ii} = \left(1 - \sum_{k=1}^{i-1} l_{ik}^2\right)^{1/2} 
   \mbox{ and } 
   l_{ji} = \left(\rho_{ij} - \sum_{k=1}^{i-1} l_{jk} l_{ik} \right)/l_{ii} .
\] 
Surprisingly, the equality in Theorem~\ref{thm:CholSums} seems to be unknown or neglected. 
The proof for Theorem~\ref{thm:CholSums} will be given in~\ref{sec:thm1proof}, and will be heavily based on the recursive arguments of the Lemma~\ref{lem:recursive}: 
\begin{lem} \label{lem:recursive}
For $i \ge 1$ and $n \ge j \ge i+1$,
\begin{equation}\label{lem:recursive_eq}
       \rhob_{i+1}^{*j}\Rb_i^{-1} \rhob_{i+1}^T  
     =
       \rhob_i^{*j}\Rb_{i-1}^{-1} (\rhob_i^{*i+1})^T  
     + \frac{ (\rho_{i,i+1} - \rhob_i^{*i+1} \Rb_{i-1}^{-1} \rhob_i^T)
              (\rho_{ij} - \rhob_i^{*j} \Rb_{i-1}^{-1} \rhob_i^T)
      }
       {1 -  \rhob_i \Rb_{i-1}^{-1} \rhob_i^T} . 
\end{equation}
\end{lem}
The proof for Lemma~\ref{lem:recursive} will be given in~\ref{sec:lemproof}.

\textbf{Remark.} Lemma~\ref{lem:recursive} is useful to illustrate the recursive nature of the computation of the part $\rhob_{i+1}^{*j}\Rb_i^{-1} \rhob_{i+1}^T$. 
As an exercise, we suggest to verify that $\rhob_2^{*j}\Rb_1^{-1} \rhob_2^T = \rho_{1j}\rho_{12}$, and then to compute $\rhob_3^{*j}\Rb_2^{-1} \rhob_3^T$ by using Eq.~(\ref{lem:recursive_eq}).
The result should be equal to $\rhob_3^{*j}\Rb_2^{-1} \rhob_3^T  = \rho_{13} \rho_{1j}  + \rho_{23(1)}\rho_{2j(1)}$ as claimed by Theorem~\ref{thm:CholSums}.

 \section{The second parametrization for Cholesky factor}\label{sec:2ndL}
The second parametrization for $\Lb$ will follow directly from~(\ref{chol:L1}) by applying Lemma~\ref{lem:ab} 
that claims an equivalence between semi-partial correlation coefficients to the difference between two successive schur-complements.
\begin{lem}\label{lem:ab}
Let us use the above notations and the positive-definiteness assumptions as before, and define $\Rb_i^{*j} \equiv  \left( \begin{array}{cc} \Rb_{i-1} & (\rhob_i^{*j})^T \\ \rhob_i^{*j} & 1 \end{array}\right)$.
Then for $j \ge i+1 \ge 3$ the difference between two successive ratios of determinants(or schur-complements) is
\begin{equation}\label{eq:lem2}
  |\Rb_i^{*j}|/|\Rb_{i-1}| - |\Rb_{i+1}^{*j}|/|\Rb_i|  
     = \left(\rho_{ij}- \rhob_i^{*j} \Rb_{i-1}^{-1} \rhob_i^T \right)^2
  |\Rb_{i-1}|/|\Rb_i|.
\end{equation}
\end{lem}
The proof for Lemma~\ref{lem:ab} will be found in~\ref{sec:lemproof}.  
Setting $s_{ij} \equiv sign(\rho_{ij(1,\ldots,i-1)})$, it is possible to write Cholesky factor~(\ref{chol:L1}) by the equivalent form
\begin{equation}\label{chol:L2}
    \left( \begin{array}{ccccc} 
             1 & 0 & 0 & \cdots & 0 \cr
            s_{12}\sqrt{1-\frac{|\Rb_2|}{1}} & \sqrt{|\Rb_2|} & 0 & \cdots & 0\cr
            s_{13}\sqrt{1-\frac{|\Rb_2^{*3}|}{1}} & s_{23}\sqrt{\frac{|\Rb_2^{*3}|}{1} - \frac{|\Rb_3|}{|\Rb_2|}} &  \sqrt{\frac{|\Rb_3|}{|\Rb_2|}} 
& \cdots & 0\cr 
\vdots & \vdots & \vdots & \ddots & 0 \cr
s_{1n}\sqrt{1-\frac{|\Rb_2^{*n}|}{1}} &  
       s_{2n}\sqrt{\frac{|\Rb_2^{*n}|}{1} - \frac{|\Rb_3^{*n}|}{|\Rb_2|}} &  
       s_{3n}\sqrt{ \frac{|\Rb_3^{*n}|}{|\Rb_2|} -  \frac{|\Rb_4^{*n}|}{|\Rb_3|}} & 
      \cdots & \sqrt{\frac{|\Rb_n|}{|\Rb_{n-1}|}  }
\end{array}\right)
\end{equation}

\subsection{The extension to nonsingular covariance matrix}\label{sec:gencov}
The Cholesky factor~(\ref{chol:L2}) can be easily extended to the general structure of nonsingular covariance when we replace each ratio $\sigma_j^2|R_i^{*j}|/|R_{i-1}|$ by its equivalent $|\Sigmab_i^{*j}|/|\Sigmab_{i-1}|$.
We summarize this result into Theorem~\ref{thm:Chlsky}.   
\begin{thm} \label{thm:Chlsky}
Let $\Sigmab_n$ be a nonsingular covariance matrix with entries $\sigma_{ij} = \sigma_i\sigma_j \rho_{ij}$. 
Let $\sigmab_i^{*j} \equiv \left(\sigma_{1j},\sigma_{2j}, \cdots, \sigma_{i-1,j}\right)$, 
set $|\Sigmab_1^{*j}|\equiv \sigma_j^2$, $|\Sigmab_0| \equiv 1$, 
and $\Sigmab_i^{*j} \equiv \left( \begin{array}{cc} \Sigmab_{i-1} & \left(\sigmab_i^{*j}\right)^T
 \cr \sigmab_i^{*j} 
 & \sigma_i^2 \end{array}\right)$.
Then, Cholesky factor $\Lb = (l_{ij})_{i,j=1}^n$ for $\Sigmab_n$ is given by $l_{ji} = 0$ (for $j < i$), and by  
\[
      l_{ji} = 
      \left\{\begin{array}{ll}
           sign(\rho_{ij(1,\ldots,i-1)}) \sqrt{|\Sigmab_i^{*j}|/|\Sigmab_{i-1}| 
                 - |\Sigmab_{i+1}^{*j}|/|\Sigmab_i|} & \mbox{ if } j >  i \\
           \sqrt{|\Sigmab_i|/|\Sigmab_{i-1}|}          &  \mbox{ if } i = j            \end{array}\right. .
\]
\end{thm}  
\textbf{Remark.} Mathematically, Theorem~\ref{thm:Chlsky} might as well describe Cholesky factor for arbitrary nonsingular symmetric matrix, when allowing the entries of $\Lb$ to be complex numbers.

\subsection{Two order conditions on the magnitudes of sub-determinants essential to a well defined $\Lb$}\label{sec:welldefL}
We start with the well known order relation between the magnitudes of successive determinants~\cite[pp. 525]{Yule1907,StuartOrdArnold2010}
\begin{equation}\label{det:order}
    1 \ge |\Rb_2| \ge |\Rb_3| \ge \cdots \ge |\Rb_{n-1}| \ge |\Rb_n| > 0.
\end{equation}
 One may alternatively see the order (\ref{det:order}) as a direct result of Eq.~(\ref{eq:schur_det}) when adding equal signs to account for $(\rhob_i = 0)$'s.
To the order~(\ref{det:order}) we will add a second order that arises from the positivity of the right-hand side of the Eq.~(\ref{eq:lem2}) and seems to be rather new. 
For $j=2,3, \ldots,n$,
\begin{equation}\label{detratio:order}
    1 \ge |\Rb_2^{*j}| \ge |\Rb_3^{*j}|/|\Rb_2|  
                   \ge \cdots \ge  
    |\Rb_{j-1}^{*j}|/|\Rb_{j-2}|\ge  
    |\Rb_j|/|\Rb_{j-1}| > 0 .
 \end{equation}
 It is possible to view the order relations (\ref{det:order}) and (\ref{detratio:order}) as posing necessary and sufficient conditions for the correlation matrix $\Rb = \Lb \Lb^T$ to be positive-definite. 
Since both order relations follow from the positive-definiteness property of $\Rb$, and on the other hand, any failure to satisfy any of the determinant ordering in (\ref{det:order}) or (\ref{detratio:order}) will lead to ill defined $\Lb$. 
In Section~\ref{sec:randomcorr} we shall show how to use the conditions (\ref{det:order}) and (\ref{detratio:order}) to generate positive-definite random correlation structures.


\section{Application I  - A simple $t$-test for linear dependence}\label{sec:ttest}  
As a first application we will establish a procedure for testing the linear dependence of a single variable upon other variables by employing the first parametrization of Cholesky factor. 
Let $\xb = (\xb_1,\cdots,\xb_p)$ be a matrix of $N$ samples from a $p$-variate random variable that is multinormally distributed. Assume that $N>p$ and let $\hat\Rb = (r_{ij})_{i,j=1}^p$ be the estimated correlation sample matrix and $\{r_{ij(1,\ldots,i-1)}\}_{i \le j}$ be the nonzero elements of Cholesky factor for $\hat\Rb$. 
Suppose that, for some $k<p$, we wish to test the linear dependence of $\xb_p$ upon $\xb_1,\xb_2,\cdots,\xb_k$. 
Consider the test 
\[ 
  H_{0k} : \rho_{1p} = \rho_{2p(1)} = \cdots = \rho_{kp(1,2,\ldots,k-1)} = 0
\] 
against the alternative 
\[
H_{1k} : \rho_{ip(1,2,\ldots,i-1)} \ne 0 \mbox{ for some } i<k.
\]
Under $H_{0k}$, the estimator $r_{kp(1,2 , \ldots ,k-1)}$ forms a simple $t$ statistic~\cite{Morrison2004}
\[
  T_{kp:N} = \frac{\sqrt{N-k} \cdot r_{kp(1,2 , \ldots ,k-1)}}
                 {\sqrt{1 - r_{kp(1,2 , \ldots ,k-1)}^2}} 
                 {H_{0k} \atop \sim } \; t_{N-k}.
\] 
The hypothesis $H_{0k}$ can be rejected, at level $\alpha$ ($0<\alpha <1$), if $|T_{kp:N}|>t_{\alpha/2,N-k}$. 
One may further establish a sequential testing procedure that searches for the largest $k$ for which $H_{0k}$ can be rejected. 

\textbf{Remark.} We leave it to the reader to verify that the null hypotheses $H_{0k}$ is equivalent to $\rho_{1p} = \rho_{2p} = \cdots = \rho_{kp} = 0$.   
 
\section{Application II - Generating realistic random correlations}\label{sec:randomcorr}
The problem of generating random correlation structures is well discussed at the literature~\cite{MarsagaliaOlkin84,Joe2006,MittelbachMatthiesenJorswieck2012}.
However, in practice, many of the suggested procedures are not so easy to apply~\cite{Holmes1991}, and when applied, some typically fail to provide a sufficient number of realistic correlation matrices~\cite{BohmHornik2014}. 
More recent algorithms for the generation of random correlations either utilize a beta distribution
~\cite{Joe2006,LewandowskiEtAl2009}, 
or employ uniform angular values~\cite{RapisardaBrigoMercurio2007,RebonatoJackel2007,MittelbachMatthiesenJorswieck2012}.
The algorithm we will suggest in this section will be considerably simple. 
It will be based on uniform values that are assigned to reflect the ratios $\{|\Rb_i^{*j}|/|\Rb_{i-1}|\}_{i\le j}$ which constitute the parametrization~(\ref{chol:L2}). 
The order of the values of $\{|\Rb_i^{*j}|/|\Rb_{i-1}|\}_{i \le j}$ will be chosen to preserve the ordering in (\ref{det:order}) and (\ref{detratio:order}), to ensure the positive-definiteness of $\Lb\Lb^T$. 
We will start by choosing $n-1$ random uniform values in $(0,1]$, that will be further assigned by their size to reflect the determinants $\{|\Rb_j|\}_{j=2}^n$, as directed by (\ref{det:order}). 
The diagonal of $\Lb$ will be constructed from the ratios $\{|\Rb_j|/|\Rb_{j-1}|\}_{j=1}^n$.  
Then, for each row of $\Lb$, $j$, an additional set of $j-2$ random uniform values will be chosen to serve as the ratios $\{|\Rb_i^{*j}|/|\Rb_{i-1}|\}_{i=2}^{j-1}$, organized to keep the order as in (\ref{detratio:order}).
The signs $s_{ij}$ will be chosen to be $(-1)^{Bernoulli(0.5)}$ and the matrix $\Lb$ will be computed according to (\ref{chol:L2}).  

\begin{enumerate}[Step 1.]
\item[\textbf{Algorithm}] \textbf{for generating realistic random correlation matrix $\Lb\Lb^T$}.
\item{\textbf{Diagonals:}} Choose $n-1$ random uniform values from the interval $(0,1]$. 
Order them in decreasing order, $U_{(2)} \ge U_{(3)} \ge \cdots \ge U_{(n)}$, 
 and set $l_{11} = 1$, $l_{22}=\sqrt{U_{(2)}}$, and $l_{jj} = \sqrt{U_{(j)}/U_{(j-1)}}$ for $j=3,\ldots,n$;
\item{\textbf{Rows:}} Repeat this step for each $j = 2,3, \ldots, n-1$. Set $U_{(1)}^{(j)}\equiv 1$ and $U_{(j)}^{(j)}\equiv l_{jj}^2$.
 For $j\ge 3$, choose $j-2$ more (additional) random uniform values $\{U_i^{(j)}\}_{i=2}^{j-1}$ inside $[l_{jj}^2,1]$ and sort them in decreasing order $U_{(2)}^{(j)} \ge \cdots \ge U_{(j-1)}^{(j)}$.
Compute $l_{ji} = \sqrt{U_{(i)}^{(j)} - U_{(i+1)}^{(j)}}$ for $i=1,2,\ldots,j-1$;
\item{\textbf{Signs:}} Choose $n(n-1)/2$ Bernoulli$(0.5)$ values $B_{ij}$'s for $j>i$, and multiply each $l_{ji}$ by $s_{ij} = (-1)^{B_{ij}}$;
\item Compute $\Rb$ by $\Lb \Lb^T$ to obtain the actual correlation structure.
\end{enumerate}

\subsection{Generating random AR(1) structures}\label{sec:AR1}
We end this paper by revealing the simple form of Cholesky factor for the $AR(1)$ structure. The AR(1) correlation matrix is defined by $\rho_{ij} = \rho^{|i-j|}$, 
and it is possible to verify that $|\Rb_n| = (1-\rho^2)^{n-1}$, $\rhob_i^{*j} = \rho^{j-i}\rhob_i$ and     
\[ 
  \rho_{ij} - \rhob_i^{*j}\Rb_{i-1}^{-1}\rhob_i^T 
  = \rho^{j-i}\left(1-\rhob_i\Rb_{i-1}^{-1}\rhob_i^T\right) 
  = \rho^{j-i} |\Rb_i|/|\Rb_{i-1}|.
\]
Cholesky factor for the AR(1) structure enjoys the simple form: 
\[
l_{ji} = \left\{\begin{array}{cc}
\rho^{j-1} & j \ge i=1 \cr
\rho^{j-i}\sqrt{1-\rho^2} & j \ge i \ge 2
\end{array}\right. 
\] 
Hence, for any choice of $\rho$, $|\rho| < 1$, it is possible to transform the standard normals $(X_i)_{i=1}^n$ into the autocorrelated normals 
\[
  Y_i = \rho^{i-1} X_1 + \sqrt{1-\rho}\sum_{k=2}^i \rho^{i-k} X_k, 
      \qquad i = 1, \ldots,n.
\] 

\section*{Acknowledgments} I am thankful to the Editor and the Reviewers for their helpful discussion and their literature suggestion. Special thanks to Professor Sen for encouraging me to submit this paper and for his resourceful suggestion to add a test of hypotheses for the application part. Thanks to Professor Speed for introducing me to Yule's paper, and to Dr. Batista for her proofreading.  

\section*{References}
\bibliographystyle{plain}

\begin{thebibliography}{99}

\biboptions{longnamesfirst,round}




\bibitem[Anderson 2003]{Anderson2003} ~Anderson, T.W. (2003). An introduction to Multivariate Statistical Analysis. Wiley, 3rd edition.


\bibitem[B\"{o}hm \& Hornik 2014]{BohmHornik2014} ~B\"{o}hm, W., and Hornik K. (2014). Generating random correlation matrices by the simple rejection method: Why it does not work. 
\emph{Statist. Probab. Lett.} \textbf{87}, 27-30.

\bibitem[Cooke et al. 2011]{CookeEtAl2011} ~Cooke, R.M., Joe, H., and Aas, K. (2011). ~Chapter 3 "Vines Arise", in ``Dependence Modeling: Vine Copula Handbook'' edited by Dorota Kurowicka and Harry Joe, World Scientific.

\bibitem[Harville 1997]{Harville1997} ~Harville, D.A. (1997). Matrix Algebra From a Statistician's Perspective, Springer.

\bibitem[Holmes 1991]{Holmes1991} ~Holmes, R.B. (1991). On Random Correlation Matrices. \emph{SIAM J. Matrix Anal. Appl.} \textbf{12}, 239-272.

\bibitem[Huber 1981]{Huber1981} ~Huber, J. (1981). Partial and Semipartial Correlation-A Vector Approach. \emph{The Two-Year College Mathematics Journal} \textbf{12}, 151-153.

\bibitem[Joe 2006]{Joe2006} ~Joe, H. (2006). ~Generating random correlation matrices based on partial correlations. \emph{J. Multivariate Anal.} \textbf{97}, 2177-2189.

\bibitem[Lewandowski et al. 2009]{LewandowskiEtAl2009} ~Lewandowski, D., Kurowicka, D., and Joe, H. (2009). Generating random correlation matrices based on vines and extended onion method. \emph{J. Multivariate Anal.} \textbf{100}, 1989-2001.

\bibitem[Marsaglia \& Olkin 1984]{MarsagaliaOlkin84} ~Marsaglia, G., and Olkin, I. (1984). Generating correlation matrices. \emph{SIAM J. Sci. Statist. Comput.} \textbf{5}, 470-475.

  \bibitem[Mittelbach et al. 2012]{MittelbachMatthiesenJorswieck2012} ~Mittelbach, M., Matthiesen, B., and Jorswieck, E. (2012). Sampling Uniformly From the Set of Positive Definite Matrices With Trace Constraint. \emph{IEEE Trans Signal Process} \textbf{60}, 2167-2179.

\bibitem[Miwa et al. 2003]{MiwaHayterKuriki2003} ~Miwa, T., Hayter, A.J., and Kuriki, S. (2003). The evaluation of general non-centered orthant probabilities. 
\emph{J. R. Stat. Soc. Ser. B}, \textbf{65}, 223-234. 

\bibitem[Moonan 1957]{Moonan57} ~Moonan, W.J. (1957).
 Linear transformation to a set of stochastically dependent normal variables. \emph{J. Am. Statist. Assoc.} \textbf{52}, 247-252.
  
\bibitem[Morrison 2004]{Morrison2004} ~Morrison, D.F. (2004). Multivariate Statistical Methods, 4th Edition.

\bibitem[Pinheiro \& Bates 1996]{PinheiroBates1996} ~Pinheiro, J.C., and Bates, D.M. (1996). Unconstrained Parameterizations for variance-covariance matrices. \emph{Statistics and Computing}, \textbf{6}, 289-296.

\bibitem[Piziak \& Odell 2007]{PiziakOdell2007} ~Piziak, R., and Odell P.L. (2007). Matrix Theory: From Generalized Inverses to Jordan Form. Chapman and Hall CRC

\bibitem[Pourahmadi 2011]{Pourahmadi2011} ~Pourahmadi, M. (2011). Covariance Estimation: The GLM and Regularization Perspectives. \emph{Statist. Sci.},\textbf{3}, 369-387.

\bibitem[Rapisarda et al. 2007]{RapisardaBrigoMercurio2007} ~Rapisarda. F., Brigo D., \& Mercurio. F. (2007). Parameterizing correlations: a geometric interpretation. \emph{IMA Journal of Management Mathematics}, \textbf{18}, 55-73.

\bibitem[Rebonato \& Jackel 2007]{RebonatoJackel2007} ~Rebonato, R., \&  Jackel, P. (2000). The most general methodology to create a valid correlation matrix for risk management and option pricing purposes. \emph{J. of Risk}, \textbf{2}, 17-27.

\bibitem[Stuart at al. 2010]{StuartOrdArnold2010} ~Stuart, A., Ord, K., and Arnold S. (2010). Kendall's Advanced Theory of Statistics, Volume 2A, Classical Inference and the Linear Model, 6th Edition.

\bibitem[Tanabe \& Sagae 1992]{TanabeSagae1992} ~Tanabe, K., and Sagae. M. (1992). An Exact Cholesky Decomposition and the Generalized Inverse of the Variance-Covariance Matrix of the Multinomial Distribution with Applications.
\emph{J. R. Stat. Soc. Ser. B}, \textbf{54}, 211-219.

\bibitem[Tong 1990]{Tong1990} ~Tong, Y.L. (1990). The multivariate normal distribution. Springer-Verlag, New York and Berlin.

\bibitem[Yule 1907]{Yule1907} ~Yule, G.U. (1907). ~On the Theory of Correlation for any Number of Variables, Treated by a New System of Notation. \emph{Proc. R. Soc. Lond. A.}, \textbf{79}, 182-193.
\end{thebibliography}

\appendix{}
\section{The proof for Theorem~\ref{thm:CholSums}}\label{sec:thm1proof}
\begin{pol2}
By mathematical induction. For $i=1$, $\rhob_2^{*j}\Rb_1^{-1} \rhob_2^T  = \rho_{1j}\rho_{12}$, 
and for $i=2$, we can use Eq.~(\ref{lem:recursive_eq}) in Lemma~\ref{lem:recursive} to get $\rhob_3^{*j}\Rb_2^{-1} \rhob_3^T = \rho_{1j}\rho_{13} + \rho_{2j(1)}\rho_{23(1)}$.
Next, suppose that $i \ge 2$ and assume that Eq.~(\ref{Q_recursive}) holds for $i$, that is, 
\begin{equation}\label{eq:ip1}
  \rhob_{i+1}^{*j}\Rb_i^{-1} \rhob_{i+1}^T  
     = \sum_{k=1}^i \rho_{k,i+1(1,\ldots ,k-1)} \cdot \rho_{kj(1,\ldots ,k-1)} ,
\end{equation}
and consider the follow-up component, $\rhob_{i+2}^{*j}\Rb_{i+1}^{-1} \rhob_{i+2}^T$, for the case of $i+1$. 
By Eq.~(\ref{lem:recursive_eq}) in Lemma~\ref{lem:recursive},
\begin{equation}\label{eq:ip2}
\begin{array}{lcl}
  \rhob_{i+2}^{*j}\Rb_{i+1}^{-1} \rhob_{i+2}^T  
    &  = & \rhob_{i+1}^{*j}\Rb_i^{-1} (\rhob_{i+1}^{*i+2})^T  
     + \frac{
         \rho_{i+1,i+2} - \rhob_{i+1}^{*i+2} \Rb_i^{-1} \rhob_{i+1}^T
       } 
       {\sqrt{1 -  \rhob_{i+1} \Rb_i^{-1} \rhob_{i+1}^T}}
        \cdot\frac{
          \rho_{i+1,j} - \rhob_{i+1}^{*j} \Rb_i^{-1} \rhob_{i+1}^T
       }
       {\sqrt{1 -  \rhob_{i+1} \Rb_i^{-1} \rhob_{i+1}^T}} \\
  &  = & \rhob_{i+1}^{*j}\Rb_i^{-1} (\rhob_{i+1}^{*i+2})^T  + 
         \rho_{i+1,i+2(1,\ldots,i)} \cdot  \rho_{i+1,j(1,\ldots,i)} .
\end{array}
\end{equation}
Since the component $\rhob_{i+1}^{*j}\Rb_i^{-1} (\rhob_{i+1}^{*i+2})^T$ has the same form as $\rhob_{i+1}^{*j}\Rb_i^{-1} \rhob_{i+1}^T$ 
with $\rhob_{i+1}^{*i+2} = (\rho_{1,i+2}, \rho_{2,i+2}, \ldots,\rho_{i,i+2})$ replacing $\rhob_{i+1} = (\rho_{1,i+1}, \rho_{2,i+1}, \ldots,\rho_{i,i+1})$, 
it is possible to rewrite $\rhob_{i+1}^{*j}\Rb_i^{-1} (\rhob_{i+1}^{*i+2})^T$ in a similar manner as was used for $\rhob_{i+1}^{*j}\Rb_i^{-1} \rhob_{i+1}^T$ in Eq. \ref{eq:ip1}
\begin{equation}\label{eq:ip3}
      \rhob_{i+1}^{*j}\Rb_i^{-1} (\rhob_{i+1}^{*i+2})^T 
    = \sum_{k=1}^i \rho_{k,i+2(1,\ldots ,k-1)} \cdot \rho_{kj(1,\ldots ,k-1)} .
\end{equation}
Combining Eq.~(\ref{eq:ip2}) with Eq.~(\ref{eq:ip3}) gives the desired Eq.~(\ref{Q_recursive}), for $i+1$: 
\[
\begin{array}{lcl}
  \rhob_{i+2}^{*j}\Rb_{i+1}^{-1} \rhob_{i+2}^T  
  &  = & \rhob_{i+1}^{*j}\Rb_i^{-1} (\rhob_{i+1}^{*i+2})^T  
     +   \rho_{i+1,i+2(1,\ldots,i)}\cdot  \rho_{i+1,j(1,\ldots,i)} \\
  &  = & \sum_{k=1}^{i+1}  \rho_{k,i+2(1,\ldots ,k-1)} \cdot \rho_{kj(1,\ldots ,k-1)} .
\end{array}
\]
\end{pol2}

\section{The proof for the two Lemmas}\label{sec:lemproof}
Lemma~\ref{lem:recursive} is for the case $l=i+1$ of the more general recursive equation:
\begin{equation}\label{eq:Rhob_jl}  
\begin{array}{l}  
 \rhob_{i+1}^{*j}\Rb_i^{-1} \left(\rhob_{i+1}^{*l}\right)^T  
   =  \rhob_i^{*j}\Rb_{i-1}^{-1}(\rhob_i^{*l})^T  
   + \frac{(\rho_{il} - \rhob_i^{*j} \Rb_{i-1}^{-1} \rhob_i^T)  
     (\rho_{ij} - \rhob_i^{*l} \Rb_{i-1}^{-1} \rhob_i^T)}{1-\rhob_i \Rb_{i-1}^{-1}\rhob_i^T}.
\end{array}
\end{equation} 
Lemma~\ref{lem:ab} is, first, by Eq.~(\ref{eq:schur_det}) 
\[
\begin{array}{ccc}
   |\Rb_i^{*j}|/|\Rb_{i-1}| - |\Rb_{i+1}^{*j}|/|\Rb_i|   
   & = & 
  1 - \rhob_i^{*j}\Rb_{i-1}^{-1} (\rhob_i^{*j})^T  
  - 1 +  \rhob_{i+1}^{*j} \Rb_i^{-1} (\rhob_{i+1}^{*j})^T ,
\end{array}
\]
and second, by applying Eq.~(\ref{eq:Rhob_jl}) for $l=j$, 
\[
\begin{array}{ccc}
   |\Rb_i^{*j}|/|\Rb_{i-1}| - |\Rb_{i+1}^{*j}|/|\Rb_i|   
   & = & 
   \frac{\left(\rho_{ij} - \rhob_i^{*j} \Rb_{i-1}^{-1}  \rhob_i^T\right)^2}{1-\rhob_i \Rb_{i-1}^{-1}\rhob_i^T} .
\end{array}
\]
\pol
By mathematical induction. 
For $i=1$, we have $\rhob_2^{*j} \Rb_1^{-1}\left(\rhob_2^{*l}\right)^T = \rho_{1l}\rho_{1j}$, and for $i=2$,
$\rhob_3^{*j} \Rb_2^{-1} \left(\rhob_3^{*l}\right)^T  = \rho_{1l} \rho_{1j} 
+ (\rho_{2l}-\rho_{12}\rho_{1l})(\rho_{2j} - \rho_{12}\rho_{1j})/(1 -\rho_{12}^2)$.
The general case of $j\ge l\ge i+1 \ge 3$ follows by representing $\Rb_i^{-1}$ 
according to the Banacheiwietz inversion formula~\citep[pp. 26]{PiziakOdell2007}, where $c_i\equiv 1-\rhob_i \Rb_{i-1}^{-1}\rhob_i^T$, 
\begin{equation}\label{Rb_i_inv}
      \Rb_i^{-1} 
       = \frac{1} {c_i}\left(\begin{array}{ccc}
       c_i \Rb_{i-1}^{-1} + \Rb_{i-1}^{-1} \rhob_i^T \rhob_i \Rb_{i-1}^{-1}  &
      - \Rb_{i-1}^{-1} \rhob_i^T \\
      - \rhob_i\Rb_{i-1}^{-1}  &  1 
      \end{array}\right) ,
\end{equation}
Finally, Eq.~(\ref{eq:Rhob_jl}) follows from writing $\rhob_{i+1}^{*j}\Rb_i^{-1} \left(\rhob_{i+1}^{*l}\right)^T$ and (\ref{Rb_i_inv})
\[  
\begin{array}{l}  
 \rhob_{i+1}^{*j}\Rb_i^{-1} \left(\rhob_{i+1}^{*l}\right)^T  
=  (\rhob_i^{*j}, \rho_{ij})\Rb_i^{-1}\left(\rhob_i^{*l}, \rho_{il}\right)^T\\
    =  \rhob_i^{*j}\Rb_{i-1}^{-1}(\rhob_i^{*l})^T  
   +(\rho_{ij} - \rhob_i^{*j} \Rb_{i-1}^{-1}  \rhob_i^T)  (\rho_{il} - \rhob_i^{*l} \Rb_{i-1}^{-1} \rhob_i^T)/c_i .
\end{array}
\]

\end{document}